\documentclass[pre,aps,twocolumn,superscriptaddress]{revtex4}
\usepackage{amsfonts}
\usepackage{amsmath}
\usepackage{graphicx}
\usepackage{amssymb}
\usepackage{txfonts}
\usepackage{braket}

\begin{document}

\title{Quantum enhancement of a single quantum battery by repeated interactions with large spins}
\author{P. Chen}
\author{T. S. Yin}
\affiliation{Physics Department of Zhejiang Sci-Tech University, Hangzhou 310018, China}
\author{Z. Q. Jiang}
\email{jiangzhongqing@163.com}
\affiliation{Physics Department of Zhejiang Sci-Tech University, Hangzhou 310018, China}
\author{G. R. Jin}
\email{grjin@zstu.edu.cn}
\affiliation{Physics Department of Zhejiang Sci-Tech University, Hangzhou 310018, China}

\begin{abstract}
A generalized collision model is developed to investigate coherent charging
a single quantum battery by repeated interactions with many-atom large spins, where collective atom operators are adopted and the battery is
modeled by a uniform energy ladder. For an initially empty battery, we derive
analytical results of the average number of excitations and hence the charging
power in the short-time limit. Our analytical results show that a faster charging and
an increased amount of the power in the coherent
protocol uniquely arise from the phase coherence of the atoms. Finally, we show that the charging power defined by the
so-called ergotropy almost follows our analytical result, due to a nearly pure state of the battery in the short-time limit.
\end{abstract}

\maketitle


\section{Introduction}

One of central goals of quantum thermodynamics is to improve the
thermodynamic processes via quantum resources and quantum operators~\cite%
{Horodecki,Niedenzu,Goold,Vinjanampathy,Campaioli}. The simplest setup to
achieve the quantum advantages in the thermodynamics is the so-called
quantum batteries (QBs)~\cite{Alicki,Hovhannisyan,Barra,Bhattacharjee},
i.e., a small quantum system that stores and provides energy. Starting from
seminal ideas developed in Ref.~\cite{Alicki}, various quantum systems have
been considered as the candidate of the QBs, including collective spins~\cite%
{Andolina99,Zhang,Peng}, interacting spin chains~\cite{Le,Juli,QZhao}, and
mechanical flywheels~\cite{Levy,SeahS}. Different to classical batteries,
the QBs explore phase coherence~\cite{Kamin} and quantum entanglement~\cite%
{Binder,CampaioliF,SantosA,RossiniD,Gyhm} as useful resources to improve the
performance of the QBs. A notable example is the Dicke-model QBs~\cite%
{Ferraro,AndolinaD,Andolina,Monsel} based on collective super-radiant
coupling in cavity and waveguide QED setups, which has been experimentally
demonstrated by using fluorescent organic molecules in a microcavity~\cite%
{QuachK}.

Recently, the QBs have been realized with a transmon qutrit~\cite{Hu} and a
solid-state qubit~\cite{Maillette,Gemme}, which clearly demonstrate the
quantum advantage at the level of a single battery~\cite%
{Seah,SalviaM,Shaghaghi}. Especially, Seah \textit{et al.}~\cite{Seah}
present a collision model to investigate the repeated charging of a single
battery by a sequence of identical qubits, where a single two-level qubit is
adopted as the charger in each interaction or collision. The collision model
describes a system that undergoes successive interactions or collisions with
the auxiliary systems~\cite{Rau,Caves,Brun,Ciccarello954}. It has been used
in various research areas, such as non-Markovian quantum dynamics~\cite%
{Ciccarello87}, quantum thermodynamics~\cite{Scarani,Strasberg}, quantum
optics~\cite{Ciccarello4}, and quantum gravity~\cite{Altamirano}. Using the
collision model, Seah \textit{et al.}~\cite{Seah} numerically show that the
phase coherence of the qubit can realize a faster charging and a larger
amount of the charging power in a comparison with that of the qubit state
without any coherence. To understand the role of the coherence, analytical
results of the charging energy and its power are necessary.

In this paper, we generalize the collision model to investigate coherent
charging of a single quantum battery by repeated interactions with many
two-level atoms (or equivalently, a large spin), where collective atomic operators
$\{\hat{J}_{\pm },\hat{J}_{z}\}$ are adopted and the battery is modeled by a
uniform energy ladder~\cite{Mitchison,Seah}. Analytical results of the
charging energy and its power are derived by considering the initially empty
battery and the short-time limit of the interaction at each charging step.
Following Ref.~\cite{Seah}, we compare the charging processes with the
atomic coherence $\langle \hat{J}_{-}\rangle \neq 0$, corresponding to the
coherent charging, and $\langle \hat{J}_{-}\rangle =0$ for the incoherent
charging. Our analytical results show that the advantage of the coherent
protocol uniquely comes from the atomic coherence. Furthermore, we show that
the amount of the charging power reaches its maximum (proportional to the
number of atoms $N_{A}$), when all the atoms prepared in a superposition
state, i.e., a coherent spin state $|\theta _{0},\phi _{0}\rangle $ with $%
\theta _{0}=\pi /2$ and $\phi _{0}=0$. Finally, we
investigate the charging power defined by the so-called ergotropy~\cite%
{Allahverdyan}. In the short-time limit, numerical results of the power show
good agreement with the analytical result, since the battery state almost
maintains in a pure state.

\section{Generalized collision model for the quantum battery}

\begin{figure}[hptb]
\begin{centering}
\includegraphics[width=1\columnwidth]{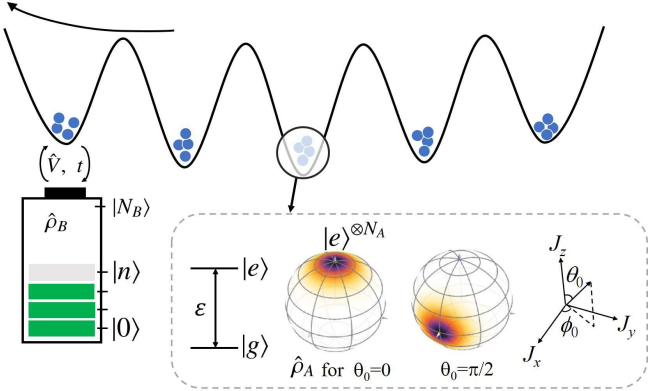}
\caption{Schematic picture of a single battery (modeled by a uniform energy ladder with $N_B+1$ levels), which undergoes successive interactions with finite $N_A$ two-level atoms confined in each lattice. The optimal atom states $\hat\rho_A$ for the incoherent and the coherent charging protocols correspond to the coherent spin states of a large spin $\ket{\theta_0, \phi_0}$, with $\theta_0=0$, $\pi/2$, as illustrated by their quasi-probability distributions on the Bloch sphere.}
\label{fig1}
\end{centering}
\end{figure}

As illustrated schematically by Fig.~\ref{fig1}, we consider a quantum
battery $\hat{\rho}_{B}$ modeled by a uniform energy ladder~\cite%
{Mitchison,Seah}, which undergoes successive interactions with identical
two-level atoms confined in several lattices. The interaction between the
battery and the atoms at each collision can be described by the Hamiltonian
\begin{equation}
\hat{H}=\hat{H}_{0}+\hat{V}=\varepsilon \hat{J}_{z}+\varepsilon \hat{n}%
+\hbar g(\hat{J}_{+}\hat{B}+\hat{J}_{-}\hat{B}^{\dag }),  \label{H}
\end{equation}%
where collective atomic operators $\hat{J}_{z}=\sum_{i=1}^{N_{A}}(|e\rangle
_{ii}\langle e|-|g\rangle _{ii}\langle g|)/2$ and $\hat{J}%
_{+}=\sum_{i=1}^{N_{A}}|e\rangle _{ii}\langle g|=(\hat{J}_{-})^{\dag }$ are
introduced to describe finite $N_A$ identical atoms, with the ground state $%
|g\rangle $ and the excited state $|e\rangle$. For the battery, one can
define the number operator $\hat{n}=\sum_{n=0}^{N_{B}}n|n\rangle \langle n|$%
, where $N_B $ denotes the highest level of the battery, and also the ladder
operators
\begin{equation}
\hat{B}=\sum_{n=1}^{N_{B}}|n-1\rangle \langle n|,\text{ \ \ }\hat{B}^{\dag
}=\sum_{n=1}^{N_{B}}|n\rangle \langle n-1|,  \label{B}
\end{equation}%
satisfying the commutation relation~\cite{BrunnerN,Erker}: $[\hat{B},\hat{B}%
^{\dag }]=|0\rangle \langle 0|-|N_{B}\rangle \langle N_{B}|$. When the
occupations of $|0\rangle $ and $|N_{B}\rangle $ are vanishing, it becomes $[%
\hat{B},\hat{B}^{\dag }]\approx 0$, corresponding to the no-boundary
condition. If we only omit the upper boundary (i.e., the occupation on $%
|N_{B}\rangle $ is vanishing), then the commutation relation is simply given
by $[\hat{B},\hat{B}^{\dag }]\approx |0\rangle \langle 0|$.

In experiments, the identical atoms (considered here as the charger) can be
realized by an ensemble of large spins at the room temperature (e.g., the
spin $j=7/2$ of cesium atoms~\cite{Fernholz,Auccaise}), or ultracold bosonic
gases in optical lattice~\cite{Widera,Riedel,Pedrozo}. On the other hand,
the energy ladder can be realized by a cavity mode with a \emph{finite}
number of energy levels~\cite{Shaghaghi,SalviaM}. In the interaction
picture, the Hamiltonian becomes $\hat{V}_{\mathrm{int}}=\exp (i\hat{H}%
_{0}t/\hbar )\hat{V}\exp (-i\hat{H}_{0}t/\hbar )=\hat{V}$, due to $[\hat{H}%
_{0},\hat{V}]=0$, which in turn gives the time evolution operator $\hat{U}%
_{\tau }=\exp [-i\tau (\hat{J}_{+}\hat{B}+\hat{J}_{-}\hat{B}^{\dag })]$,
where $\tau =gt$ and $t$ denotes the interaction time at each charging step.
Starting from an initial state $\hat{\rho}_{B}(0)$, the battery state at
the $k$th collision becomes
\begin{equation}
\hat{\rho}_{B}(k)=\mathrm{Tr}_{A}[\hat{U}_{\tau }\hat{\rho}_{B}(k-1)\otimes
\hat{\rho}_{A}\hat{U}_{\tau }^{\dag }],  \label{rhobk}
\end{equation}%
where $\mathrm{Tr}_{A}(\cdots )$ denotes the trace over the atom states.
Similar to Ref.~\cite{Seah}, we assume that the atom states are identical
for all the collisions (i.e., $\hat{\rho}_{A}$ is independent on $k$). In
the limit $\tau \rightarrow 0$, the time evolution operator can be
approximated as $\hat{U}_{\tau }\approx 1-i\tau (\hat{J}_{+}\hat{B}+\hat{J}%
_{-}\hat{B}^{\dag })$, and therefore
\begin{equation}
\hat{\rho}_{B}(k)\approx \hat{D}^{\dag }(k\alpha )\hat{\rho}_{B}(0)\hat{D}%
(k\alpha ),  \label{rho0}
\end{equation}%
where $\hat{D}(\alpha )=\exp (\alpha \hat{B}^{\dag }-\alpha ^{\ast }\hat{B})$
and $\alpha =i\tau \langle \hat{J}_{-}\rangle $, with $\langle (\cdots
)\rangle =\mathrm{Tr}_{A}[\hat{\rho}_{A}(\cdots )]$. Similar result of Eq.~(%
\ref{rho0}) has been obtained by Ref.~\cite{SalviaM} (see also the Appendix A).

Using the probability distribution $P_{n}(k)=\langle n|\hat{\rho}%
_{B}(k)|n\rangle $, one can define the mean number of excitations and the
mean energy~\cite{Seah}:
\begin{equation}
\bar{n}_{k}=\sum_{n=0}^{N_{B}}nP_{n}(k)\text{, \ \ \ }E(k)=\varepsilon \bar{n%
}_{k},  \label{en}
\end{equation}%
where $\varepsilon $ denotes the energy spacing of the battery. Following
Ref.~\cite{Seah}, we first consider the no-boundary condition (i.e., the
occupations on $|0\rangle $ and $|N_{B}\rangle $ are vanishing), which
allows us to obtain a recursion relation (see the Appendix~A),
\begin{equation}
\bar{n}_{k}=\bar{n}_{k-1}+v+\mathrm{Im}\left( \Omega \beta _{k-1}^{\ast
}\right) ,  \label{nk}
\end{equation}%
where $v=2\sin ^{2}(\tau )\langle \hat{J}_{z}\rangle $, $\Omega =\sin (2\tau
)\langle \hat{J}_{-}\rangle $, and $\beta _{k}=\mathrm{Tr}_{B}[\hat{\rho}%
_{B}(k)\hat{B}]$. Without any boundary, we have $\beta_{k}=\beta _{0}$, and therefore $\bar{n}_{k}=\bar{n}_{0}+k[v+\mathrm{Im}%
(\Omega \beta _{0}^{\ast })]$, where Eq.~(\ref{nk}) has iterated for $k$
times. One can easily find that the mean number of excitations $\bar{n}_{k}$
and hence the mean energy of the battery $E(k)$ grow linearly with respect
to the number of charging steps $k$~\cite{Seah}. When $\bar{n}_{k}\approx
N_{B}$, the battery can be regarded as being fully charged.

Next we focus on the short-time limit (i.e., $\tau \rightarrow 0$) to obtain
$v\approx2\tau ^{2}\langle \hat{J}_{z}\rangle \sim 0$ and $%
\Omega\approx2\tau \langle \hat{J}_{-}\rangle =-2i\alpha $, which yield
\begin{equation}
\bar{n}_{k}\approx \bar{n}_{k-1}-(\alpha \beta _{k-1}^{\ast }+\alpha ^{\ast
}\beta _{k-1}),  \label{nklimit}
\end{equation}%
where the lower boundary $|0\rangle $ has been taken into account (see the
Appendix~B), as
\begin{equation}
\beta _{k}=\beta _{0}-\alpha \sum_{k^{\prime }=0}^{k-1}\langle 0|\hat{\rho}%
_{B}(k^{\prime })|0\rangle.  \label{beta2}
\end{equation}%
Note that the above recursion relations of the average number of
excitations, i.e., Eqs.~(\ref{nk}) and (\ref{nklimit}) are independent on
any specific form of $\hat{\rho}_{A}$, and even free from the initial state
of $\hat{\rho}_{B}(0)$. Next, we consider a specific form of $\hat{\rho}_{A}$
and extend the single-particle case (i.e., $N_{A}=1$)~\cite{Seah} into the
many-particle case, for which the total charging process becomes more faster
and the amount of charging power can be increased, dependent on $N_{A}$.

\section{Collectively Coherent charging}

In Ref.~\cite{Seah}, Seah \textit{et al.} consider the battery charged by a
sequence of single atom (i.e., $N_{A}=1$), with the atom state~\cite%
{Radcliffe,Arecchi,ZhangD,Kitagawa,Jin,Ji}
\begin{eqnarray}
\hat{\rho}_{A} &=&\cos ^{2}\left( \frac{\theta _{0}}{2}\right) |e\rangle
\langle e|+\sin ^{2}\left( \frac{\theta _{0}}{2}\right) |g\rangle \langle g|
\notag \\
&&+c\left( \frac{\sin (\theta _{0})}{2}e^{-i\phi _{0}}|e\rangle \langle g|+%
\mathrm{H.c.}\right) ,
\end{eqnarray}%
where $\theta _{0}$ and $\phi _{0}$ determine the population imbalance and
the relative phase between the two states~\cite{Kitagawa,Jin,Ji}, as shown
in Fig.~\ref{fig1}. The parameter $c\in \lbrack 0,1]$ is added artificially
to distinguish the two opposing protocols: the incoherent charging ($c=0$)
and the coherent charging ($c=1$). The spin-$1/2$ can be mapped into a large
spin system with $j=N_A/2$, for which the atom state becomes
\begin{equation}
\hat{\rho}_{A}=\sum_{m}\rho _{m,m}|j,m\rangle \langle j,m|+c\sum_{m\neq
m^{\prime }}\rho _{m,m^{\prime }}|j,m\rangle \langle j,m^{\prime }|,
\label{css}
\end{equation}%
where $\{|j,m\rangle\}$ denote eigenstates of $\hat{J}_{z}$ and\ $\rho
_{m,m^{\prime }}=d_{m}d_{m^{\prime }}^{\ast }$, with
\begin{equation*}
d_{m}=\binom{2j}{j+m}^{1/2}\cos ^{j+m}\left( \frac{\theta _{0}}{2}\right)
\sin ^{j-m}\left( \frac{\theta _{0}}{2}\right) e^{i(j-m)\phi _{0}},
\end{equation*}%
and $\binom{n}{m}=\frac{n!}{m!(n-m)!}$. For the coherent charging scheme
(i.e., $c=1$), $\hat{\rho}_{A}$ becomes a coherent spin state $|\theta
_{0},\phi _{0}\rangle $~\cite{Radcliffe,Arecchi,ZhangD,Kitagawa,Jin}, which
gives the population imbalance and the phase coherence determined by $%
\langle \hat{J}_{z}\rangle =j\cos (\theta _{0})$ and $\langle \hat{J}%
_{-}\rangle=cj\sin (\theta _{0})\exp (-i\phi _{0})$~\cite{Jin,Ji},
respectively. Hereafter, we choose the azimuthal angle $\phi _{0}=0$ and
therefore,
\begin{equation}
v=2j\cos (\theta _{0})\sin ^{2}(\tau )\text{, \ \ \ \ \ }\Omega =cj\sin
(\theta _{0})\sin (2\tau ),  \label{vOmega}
\end{equation}%
where $j=N_{A}/2$, and $c=\Omega =0$ for the incoherent protocol.

\begin{figure}[hptb]
\begin{centering}
\includegraphics[width=1\columnwidth]{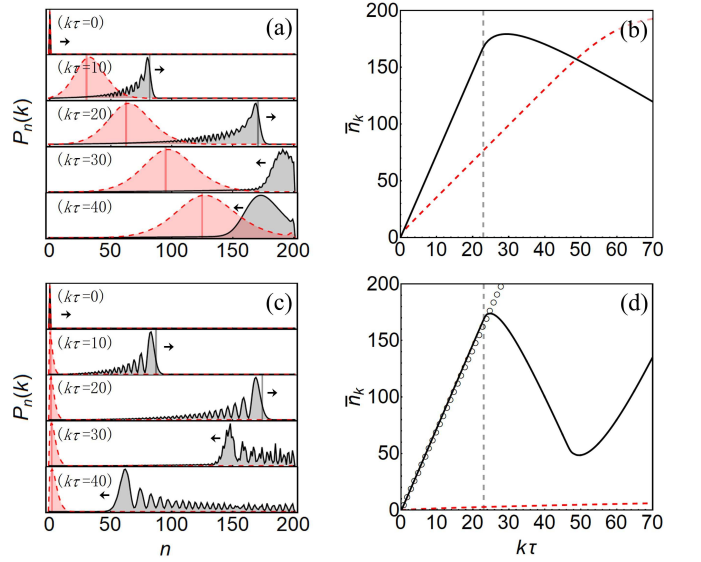}
\caption{Probability distributions of $\hat{\rho}%
_{B}(k)$ (left panel) and average number of the excitations (right panel), for different values of the charging time $\tau=\pi/4$ (in a and b) and $0.01$ (c and d). Solid (red dashed) lines correspond to the coherent (incoherent) charging protocol. All curves in the left panel are rescaled to their associated maxima. Vertical lines: location of the peak $n\approx (v+\Omega)k$ (left panel) and $k_\mathrm{est}\tau$ (right panel), determined by Eqs.~(\ref{n bound}) and (\ref{kest}), respectively. Circles in (d): analytical result of $\bar n_k$, given by Eq.~(\ref{nk_ana}). Parameters: $N_A=10$, $N_B=200$, and $\theta_0=\pi/3$.}
\label{fig2}
\end{centering}
\end{figure}

We first consider the incoherent charging process to an initially empty
battery $\hat{\rho}_{B}(0)=|0\rangle \langle 0|$, using $\hat{\rho}_{A}$
with $c=0$ and $\theta _{0}=\pi /3$. The red dashed lines of Fig.~\ref{fig2}%
(a) show the probability distributions $P_{n}(k)$ against $n$ for different
charging times $k\tau $, where $\tau =\pi /4$ is fixed~\cite{Seah}. One can
find that the probability distribution tends to a Gaussian as $k$ increases.
Therefore, from Eq.~(\ref{nk}), the peak of $P_{n}(k)$ appears at
\begin{equation}
n=\bar{n}_{k}\approx vk+\Omega \sum_{k^{\prime }=0}^{k-1}\mathrm{Im}(\beta
_{k^{\prime }}^{\ast })\sim (v+\Omega )k,  \label{n bound}
\end{equation}%
where the last result holds when $\mathrm{Im}(\beta _{k})\sim 1$ (see the
Appendix B). From Fig.~\ref{fig2}(a), one can see the locations of the peaks
$n\approx (v+\Omega )k$, as indicated by the vertical lines. When the peak
approaches to the highest level (i.e., $n\approx N_{B}$), the battery can be
regarded as being fully charged and the number of charging steps needed is
given by \cite{Seah}:
\begin{equation}
k_{\mathrm{est}}\approx \mathrm{Ceiling}\left[ \frac{N_{B}}{v+\Omega }\right]
,  \label{kest}
\end{equation}%
where $\mathrm{Ceiling}[x]$ gives the smallest integer greater than or equal
to $x$. In Fig.~\ref{fig2}(b), we show the mean value of the excitations $%
\bar{n}_{k}$ against the total charging time $k\tau $. As shown by the red
dashed line, $\bar{n}_{k}$ monotonically increases from $0$ to its maximal
value $N_{B}$. Similar result has been observed by considering the
single-qubit case of $\hat{\rho}_{A}$~\cite{Seah}.

For the single-qubit case~\cite{Seah}, it has been shown that the number of
collisions $k_{\mathrm{est}}\approx 800$, corresponding to the total
charging time $k_{\mathrm{est}}\tau \approx 628$. For the coherent charging
process (i.e., $c=1$), it is about $k_{\mathrm{est}}=292$~\cite{Seah} and
hence the total charging time $k_{\mathrm{est}}\tau $ $\approx 229$, which
is shorter than that of the incoherent case by $2.74$ times. To understand
the advantage of the coherent protocol, one can note that the number of
collisions $k_{\mathrm{est}}$ can be reduced due to $\Omega \neq 0$ (it is
maximized for $\tau =\pi/4$ and is vanishing for the incoherent case), and
therefore leads to a faster charging. However, when $k>k_{\mathrm{est}}$,
the coherent protocol losses its advantage, due to a decay of $\bar{n}_{k}$.
This is because the probability distribution $P_{n}(k)$ is reflected by the
upper boundary ($n=N_{B}$)~\cite{Seah}, which reduces $\bar{n}_{k}$ and
hence the energy of the battery $E(k)$. Such a phenomenon also occurs for
the many-particle case.

For the many-particle case, e.g., $N_{A}=10$, the charging time can be
further reduced for both the incoherent and the coherent protocols. As shown
by Fig.~\ref{fig2}(b), the incoherent scheme requires $k_{\mathrm{est}}\tau
\approx 63$ (the red dashed line) to fully charging the battery. For the
coherent case (the solid line), the total charging time is about $k_{\mathrm{%
est}}\tau $ $\approx 23$. Indeed, the number of charging steps $k_{\mathrm{%
est}}$ is reduced by $N_{A}$ times since both $v$ and $\Omega $ are
proportional to $j$ ($=N_{A}/2$), as Eqs.~(\ref{vOmega}) and (\ref{kest}).
Therefore, a more faster charging can be realized by using the many-particle
$\hat{\rho}_{A}$.

In Fig.~\ref{fig2}(c) and (d), we further consider the above two charging
processes in the short-time limit (e.g., $\tau=10^{-2}$). For this case, the
coherent protocol always outperforms the incoherent one within the total
charging time $k\tau \in \lbrack 0$, $70]$. For the incoherent case, both $%
P_{n}(k)$ and $\bar{n}_{k}$ increase very slowly (the red dashed lines).
Indeed, the battery is fully charged when $k_{\mathrm{est}}\tau \approx
4\times 10^3$. In contrast, the coherent scheme significantly reduces the
total charging time. One can see that $\bar{n}_{k}$ increases to its maximum
at $k_{\mathrm{est}}\tau \approx 23$, with a very small value of interaction
time $\tau =10^{-2} $. This result is somewhat counter-intuitive.

To understand the above result, we derive analytical results of $\bar{n}_{k}$
and hence the charging power in the short-time limit (i.e., $\tau \ll 1$).
According to Eqs.~(\ref{nklimit}) and (\ref{beta2}), $\bar{n}_{k}$ depends on $\beta _{k}$ and
also $\langle 0|\hat{\rho}_{B}(k)|0\rangle \approx |\langle 0|\hat{D}%
(k\alpha )|0\rangle |^{2}$, where
\begin{equation}
\langle 0|\hat{D}(k\alpha )|0\rangle =\sum_{l=0}^{\infty }\frac{k^{l}}{l!}%
\langle 0|(\alpha \hat{B}^{\dag }-\alpha ^{\ast }\hat{B})^{l}|0\rangle ,
\label{0D0}
\end{equation}%
with $\alpha =i\tau \langle \hat{J}_{-}\rangle $. Performing a series
expansion over $\langle 0|(\alpha \hat{B}^{\dag }-\alpha ^{\ast }\hat{B}%
)^{l}|0\rangle $, one can see that it is vanishing for odd $l$ (see the
Appendix B). As inspired by Fig.~\ref{fig3} (a), for even $l=2n$, we obtain $%
\langle 0|(\alpha \hat{B}^{\dag }-\alpha ^{\ast }\hat{B})^{2n}|0\rangle
=(-1)^{n}|\alpha |^{2n}C_{n}$, where $C_{n}=\frac{1}{n+1}\binom{2n}{n}$
denotes the Catalan number~\cite{Hilton,Stanley}, and therefore
\begin{equation}
\langle 0|\hat{D}(k\alpha )|0\rangle =\sum_{n=0}^{\infty }\frac{%
(-1)^{n}(k|\alpha |)^{2n}C_{n}}{(2n)!}=\frac{J_{1}(2k|\alpha |)}{k|\alpha |}.
\label{ka00}
\end{equation}%
Here $J_{1}(x)$ denotes the first-order Bessel function of the first kind.
Now Eq.~(\ref{beta2}) becomes
\begin{equation}
\beta _{k}\approx \beta _{0}-\alpha \sum_{k^{\prime }=0}^{k-1}\left( \frac{%
J_{1}(2k^{\prime }|\alpha |)}{k^{\prime }|\alpha |}\right) ^{2},
\label{beta_k}
\end{equation}%
where $\beta _{0}=\mathrm{Tr}_{B}[\hat{\rho}_{B}(0)\hat{B}]=0$. From Eq.~(%
\ref{nklimit}), we further obtain (see the Appendix B)
\begin{equation}
\bar{n}_{k}\approx 2\sum_{k^{\prime }=0}^{k-2}(k-1-k^{\prime })\left( \frac{%
J_{1}(2k^{\prime }|\alpha |)}{k^{\prime }}\right) ^{2},  \label{nk_ana}
\end{equation}%
where $\bar{n}_{0}=0$. For a large enough $k$ (so that $k-2\approx k$), the
sum over $k^{\prime }$ can be replaced by an integral
\begin{equation}
\bar{n}_{k}\approx 2\int_{0}^{k}(k-k^{\prime })\left( \frac{J_{1}(2k^{\prime
}|\alpha |)}{k^{\prime }}\right) ^{2}dk^{\prime }=2xf(x),  \label{nk_integ}
\end{equation}%
where $x=k|\alpha |=k\tau \langle \hat{J}_{-}\rangle $, with $\langle \hat{J}%
_{-}\rangle =j\sin (\theta _{0})$, and
\begin{equation}
f(x)=\frac{1}{x}\int_{0}^{x}\left( x-x^{\prime }\right) \left[ \frac{%
J_{1}(2x^{\prime })}{x^{\prime }}\right] ^{2}dx^{\prime }.  \label{fx}
\end{equation}%
From Fig.~\ref{fig3}(b), one can see that $f(x)$ is a monotonic function,
which increases from $0$ to its asymptotic value $f(\infty
)=8/(3\pi)\approx0.85$. When $x=30$, $f(x)$ approaches to $f(\infty)$.

\begin{figure}[hptb]
\begin{centering}
\includegraphics[width=1\columnwidth]{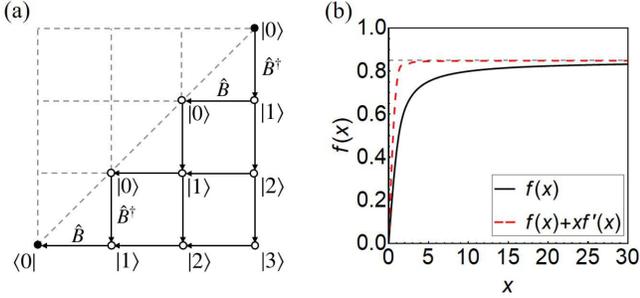}
\caption{(a) All possible paths for $\langle 0|(\alpha ^{\ast
}\hat{B}-\alpha \hat{B}^{\dag })^{2n}|0\rangle\neq0$, with the number of paths given by the Catalan number $C_n$. (b) The monotonic function $f(x)$ (solid), defined by Eq.~(\ref{fx}), and $f(x)+xf'(x)$ (red dashed). In (a), the number of paths $C_n=5$ for $n=3$. Horizontal line in (b): $8/(3\pi)$. }
\label{fig3}
\end{centering}
\end{figure}

Note that Eqs.~(\ref{nk_ana}) and (\ref{nk_integ}) are valid for $\hat{\rho}%
_{B}(0)=|0\rangle\langle0|$ and $k\tau\leq k_{\mathrm{est}}\tau$, where the
charging time per collision $\tau\ll 1$. In the short-time limit, from Eq.~(%
\ref{kest}), we obtain $k_{\mathrm{est}}\tau \approx \tau \mathrm{Ceiling}%
[N_{B}/\Omega ]\sim N_{B}/N_{A}$, provided $\theta _{0}=\pi /2$. As depicted
by Fig.~\ref{fig2}(d), our analytical results (the circles) shows a good
agreement with the numerical result of $\bar{n}_{k}$, as long as $k\tau\leq
k_{\mathrm{est}}\tau \approx 23$. Furthermore, Eq.~(\ref{nk_ana}) at $k=k_{%
\mathrm{est}}$ gives $\bar{n}_{k}\approx 0.82N_{B}$, coincident quite well
with the numerical result $0.87N_{B}$. Numerical results of $\bar{n}_{k}$
decreases after $k\tau \gtrsim k_{\mathrm{est}}\tau $, which can not be
predicted by our analytical result.

\section{Charging power}

The performance of the battery can be quantified by the charging power
\begin{equation}
P=\frac{E(k)-E(0)}{kt}=g\varepsilon \left( \frac{\bar{n}_{k}-\bar{n}_{0}}{%
k\tau }\right) ,  \label{p}
\end{equation}%
where $E(0)=\varepsilon \bar{n}_{0}$ is the mean energy of the initially
uncharged battery. For the fully empty battery, $\hat{\rho}%
_{B}(0)=|0\rangle\langle0|$, we have $E(0)=\bar{n}_{0}=0$. Using Eq.~(\ref{n
bound}), one can obtain an approximate upper bound of the power,
\begin{equation}
P\approx \frac{g\varepsilon }{\tau }(v+\Omega )=g\varepsilon N_{A}\frac{\sin
(\tau )}{\tau }\sin (\tau +\theta _{0})\leq g\varepsilon N_{A},
\label{p_bound}
\end{equation}%
where $v$ and $\Omega $ are given in Eq.~(\ref{vOmega}). The first result
comes from Eq.~(\ref{n bound}) when $\mathrm{Im}(\beta _{k})\sim 1$, valid
for the no-boundary condition. The second inequality is a natural result of
Eq.~(\ref{vOmega}) with $c=1$. One can easily find that maximum of the power
can be reached at $\theta _{0}=\pi /2-\tau $. Therefore, in the limit $\tau
\rightarrow 0$ (i.e., $\theta _{0}\approx \pi/2$), the power is possible to
reach its upper bound $g\varepsilon N_{A}$.

To reach the upper bound, we now consider the coherent charging of the
initially empty battery (i.e., $c=1$ and $\hat{\rho}_{B}(0)=|0\rangle \langle
0|$), using an optimal atom state $|\pi /2,0\rangle $. In\ the short-time
limit $\tau \rightarrow 0$, Fig.~\ref{fig2}(d) suggests a large enough
number of the atom-battery interactions (i.e., $k\rightarrow \infty $, but
with a finite total charging time $k\tau $). Using Eq.~(\ref{nk_integ}), we
obtain the analytical result of the power
\begin{equation}
P_{\mathrm{coh}}\approx \frac{2g\varepsilon }{k\tau }xf(x)=2g\varepsilon
\langle \hat{J}_{-}\rangle f(x),  \label{pcoh}
\end{equation}%
where $x=k\tau \langle \hat{J}_{-}\rangle $, with $\langle \hat{J}%
_{-}\rangle =j\sin (\theta _{0})$. The optimal atom state $|\theta
_{0},0\rangle $ with $\theta _{0}=\pi /2$ can be obtained from the following
equation
\begin{equation}
0=\frac{\partial P_{\mathrm{coh}}}{\partial \theta _{0}}\propto \frac{%
\partial x}{\partial \theta _{0}}\left[ f(x)+x\frac{\partial f(x)}{\partial x%
}\right] ,
\end{equation}%
or equivalently, $0=\partial x/\partial \theta _{0}\propto \partial \langle
\hat{J}_{-}\rangle /\partial \theta _{0}=j\cos (\theta _{0})$, i.e., $\theta
_{0}=\pi /2$. With the optimal state, the charging
power $P_{\mathrm{coh}}$ can reach its maximum at $k_{\mathrm{est}}\tau \sim
N_{B}/N_{A}$, with
\begin{equation}
P_{\mathrm{coh,\max }}\approx N_{A}g\varepsilon f(\infty )=\frac{8}{3\pi }%
N_{A}g\varepsilon ,  \label{p_cohmax}
\end{equation}%
where $f(\infty )=8/(3\pi )\approx 0.85$, as shown by Fig.~\ref{fig3} (b).
In the single-particle picture, the optimal atom state can be rewritten as a
direct product $|\pi /2,0\rangle \varpropto (|e\rangle +|g\rangle )^{\otimes
N_{A}}$, corresponding to all the spins pointed to the $\hat{J}_{x}$ axis,
as depicted in Fig.~\ref{fig1}. The coherent spin state with $\theta
_{0}=\pi /2$ has been prepared in the large spin system at the room
temperature~\cite{Fernholz,Auccaise} and the ultracold bosonic gases in
optical lattice~\cite{Widera,Riedel,Pedrozo}.

\begin{figure}[hptb]
\begin{centering}
\includegraphics[width=1\columnwidth]{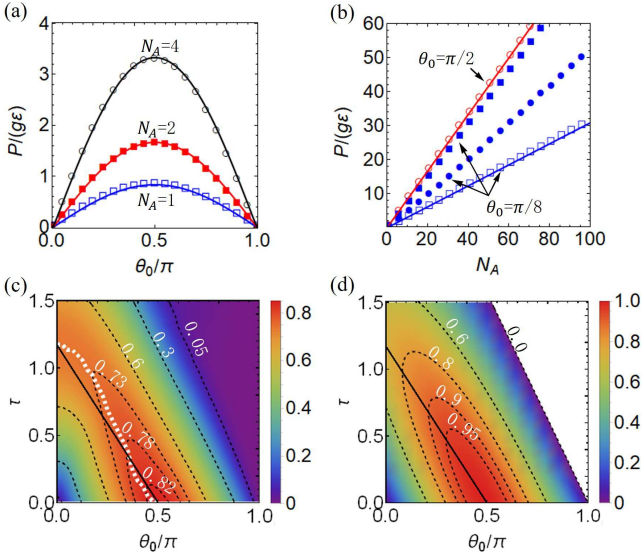}
\caption{(a) Scaled power $P/(g\varepsilon)$ against $\theta_0$ for a fixed $\tau=0.01$ and different numbers of the atoms, (b) the charging power against $N_A$ for $\theta_0=\pi/8$ and different values of $\tau$, (c) and (d) numerical and analytical results of $P/(g\varepsilon N_A)$ as functions of $\theta_0$ and $\tau$. All the numerical results are obtained at $k\tau=60/N_A$. In (a): from top to bottom, $N_A=4$ (open circles), $2$ (squares), $1$ (open squares), coincident with the curves for Eq.~(\ref{pcoh}). In (b): from top to bottom, $\tau=0.88$ (squares), $0.3$ (circles), $0.01$ (open squares), to comparing with the optimal result (the open circles for $\theta_0=\pi/2$ and $\tau=0.01$), and the curves also from Eq.~(\ref{pcoh}). The white dashed line of (c): local maximum of the power at $k\tau=60/N_A$ for $N_A=10$ and different values of $(\theta_0, \tau)$, to comparing with its analytical result $\tau=1.17(1-2\theta_0/\pi)$ (solid line), obtained from the second result of Eq.~(\ref{p_bound}). All for $N_B=200$.}
\label{fig4}
\end{centering}
\end{figure}

The coherent charging scheme is robust to imperfections in preparing the
atomic states. As shown in Fig.~\ref{fig4}(a), one can see that $P_{\mathrm{%
coh}}$ varies smoothly with $\theta _{0}$ for different values of $N_{A}$. No peak or dip at $\theta_0\sim\pi/2$ means that the coherent scheme
works well for a sub-optimal atom state. When $\theta_0$ largely departures from $%
\pi/2$ (e.g., $\theta_0=\pi/8$), the squares of Fig.~\ref{fig4}(b) show that the power $P_{%
\mathrm{coh,\max}}\approx 0.85g\varepsilon N_{A}$ can almost maintain by
taking a relatively larger charging time $\tau$, which has not been investigated in Ref.~\cite%
{Seah}.

To confirm the above results, we consider the coherent charging to the
initially empty battery (i.e., $c=1$ and $\hat{\rho}_{B}(0)=|0\rangle \langle
0|$). In Fig.~\ref{fig4}, we choose a fixed charging time $k\tau =60/N_{A}$ to calculate numerical
results of $\bar{n}_{k}$ and hence the power $P$, which depend on the choices of $\theta _{0}$ and $\tau$. For
given $N_{A}=1$ (the open squares), $2$ (the squares), $4$ (the open circles), in Fig.~\ref{fig4}(a), we take $\tau =0.01$
(i.e., from bottom to top, $k=6000$, $3000$, and $1500$) and show the scaled
power as a function of $\theta _{0}$. As $\tau\ll 1$, one can see that our
analytical results work well (the curves) and the maximum of the power appears
at $\theta _{0}=\pi /2$. When $\theta_0$ largely departures from $\pi/2$, e.g., $\theta
_{0}=\pi /8$ in Fig.~\ref{fig4}(b), we show the power as a function of $%
N_{A}$ for $\tau =0.01$ (the open squares), $0.3$ (the circles), $0.88$
(the squares). From the squares, one can see that the optimal result $P_{\mathrm{coh,\max }}$ that
obtained for $\theta _{0}=\pi /2$ and $\tau\ll1$ (the open circles) almost maintains by
choosing a relatively larger charging time $\tau$.

In Fig.~\ref{fig4}(c), we take $N_{A}=10$ and show the
scaled power $P/(g\varepsilon N_{A})$ against $\theta _{0}$ and $\tau $.
When $\tau \ll 1$, the optimal atom state corresponds to $\theta _{0}=\pi /2$, as expected. While for each value of $\theta _{0}<\pi /2$, there exists an optimal value of the
charging time $\tau$ (the white dashed line). In Fig.~\ref{fig4}(d), we
show the scaled power using the second result of Eq.~(\ref{p_bound}), where
the white background region indicates $P\propto \sin (\tau +\theta _{0})<0$.
This is unphysical. Analytically, we show that the maximum of the power can
be obtained for $\tau =1.17(1-2\theta _{0}/\pi )$ (the solid line),
coincident quite well with the white dashed line of Fig.~\ref{fig4}(c). When
$\theta _{0}=\pi /8$, we have $\tau =0.88$, as depicted by the squares of
Fig.~\ref{fig4}(b). As $\theta _{0}\rightarrow 0$, the maximum of the power appears
at $\tau =1.17$, which can be understood by considering the incoherent
charging protocol.

For the incoherent charging case (i.e., $c=\Omega =0$), Eq.~(\ref{nk})
becomes $\bar{n}_{k}\approx kv$, and therefore
\begin{equation}
P_{\mathrm{inc}}\approx 2g\varepsilon \langle \hat{J}_{z}\rangle \frac{\sin
^{2}(\tau )}{\tau },  \label{p_inc}
\end{equation}%
where $\langle \hat{J}_{z}\rangle =j\cos (\theta _{0})$. One can easily find
that an optimal value of $P_{\mathrm{inc}}$ can be obtained for $\theta
_{0}=0$, corresponding to an optimal atom state $|e\rangle
^{\otimes N_{A}}$, as shown by Fig.~\ref{fig1}. Furthermore, $P_{\mathrm{inc}%
}$ reaches its maximum at a finite $\tau $, determined by%
\begin{equation}
0=\left. \frac{\partial P_{\mathrm{inc}}}{\partial \tau }\right\vert _{\tau
_{0}}\propto \frac{\sin ^{2}\tau _{0}}{\tau _{0}^{2}}\left( 2\tau _{0}\cot
\tau _{0}-1\right) ,
\end{equation}%
or equivalently, $\tau _{0}\cot \tau _{0}=1/2$. This is a transcendental
equation with one of the roots $\tau _{0}\approx 1.17$~\cite{Seah}, which
gives $P_{\mathrm{inc,\max }}\approx 0.72g\varepsilon N_{A}$. Note that $P_{%
\mathrm{inc,\max }}$ is the achievable power attained from the incoherent
charging protocol, as depicted by the top edge of the gray area in Fig.~\ref%
{fig5}(a) and (b). Using the optimal state with $\theta _{0}=0$, Eq.~(\ref%
{kest}) indicates that fully charging the battery with the incoherent
protocol requires the time $k_{\mathrm{est}}\tau _{0}\approx \tau _{0}%
\mathrm{Ceiling}[N_{B}/v]\sim 1.38N_{B}/N_{A}$, larger than that of the
coherent charging protocol ($k_{\mathrm{est}}\tau \sim N_{B}/N_{A}$ for $%
\theta _{0}=\pi /2$).

\begin{figure}[hptb]
\begin{centering}
\includegraphics[width=1\columnwidth]{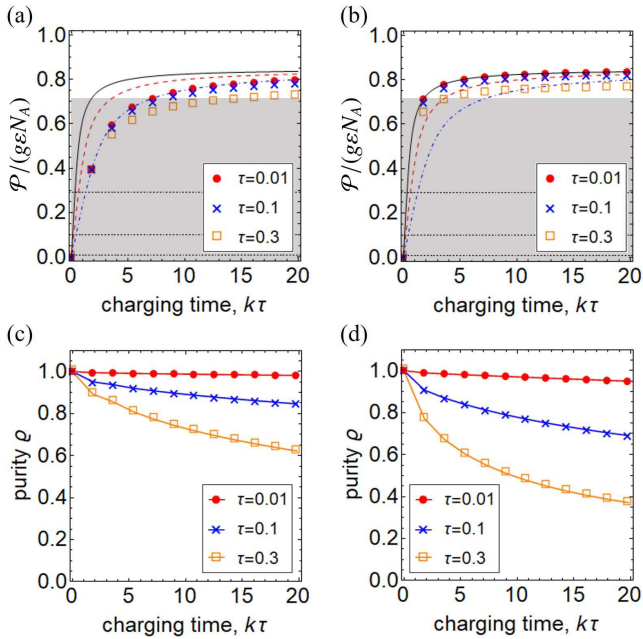}
\caption{With a given $\tau=0.01$, $0.1$, and $0.3$, the scaled power $\mathcal{P}/(g\varepsilon N_A)$ defined by the ergotropy and the purity of $\hat\rho_B(k)$ for $N_A=1$ (left panel) and $4$ (right panel). The curves in (a) and (b): the scaled power $P_{\mathrm{coh}}/(g\varepsilon N_A)$ for $N_A=1$ (dot-dashed), $2$ (dashed), and $4$ (solid), obtained from Eq.~(\ref{pcoh}) with $\theta_0=\pi/2$. The horizontal lines in (a) and (b): the power of the incoherent charging scheme $P_\mathrm{inc}/(g\varepsilon N_A)$ for different values of $\tau=0.01$, $0.1$, and $0.3$ (from bottom to top). The top edge of the gray area: the maximum power of the incoherent charging protocol for $\tau=1.17$, i.e., $P_{\mathrm{inc,\max }}/(g\varepsilon N_A)\simeq 0.72$. All for $N_B=200$.}
\label{fig5}
\end{centering}
\end{figure}

Comparing Eqs.~(\ref{pcoh}) and (\ref{p_inc}), one can easily find that the
coherent protocol depends on the atomic coherence $\langle \hat{J}%
_{-}\rangle $, and the power varies with the time $x=k\tau \langle \hat{J}%
_{-}\rangle $. For the incoherent one, however, the power is a function of $%
\tau $, independent on the atomic coherence. Our analytical results show that
the advantages of the coherent protocol in the charging time and that of the
charging power \emph{uniquely} arise from the phase coherence of the atoms.
For the coherent spin state with $\theta _{0}=\pi /2$, the coherence becomes
maximum and the charging power $P_{\mathrm{coh,\max }}\approx
0.85g\varepsilon N_{A}$ can be obtained for the charging time $\tau
\rightarrow 0$, which significantly reduces the role of noise during each
charging step. Furthermore, with a finite total charging time $k_{\mathrm{est%
}}\tau \sim N_{B}/N_{A}$, one can reduce the number of charging steps $k_{%
\mathrm{est}}$ by using large $N_{A}$, which is inaccessible from Ref.~\cite%
{Seah}.

Finally, it should be mentioned that the useful energy of $\hat{\rho}_{B}(k)$
that can be extracted is given by the so-called ergotropy~\cite{Allahverdyan}:
\begin{equation}
\mathcal{E}_{B}(k)=E(k)-\mathrm{Tr}_{B}[\hat{\sigma}_{B}(k)\hat{H}_{B}],
\label{ergotropy}
\end{equation}%
where $\hat{H}_{B}=\varepsilon \hat{n}$ is the free Hamiltonian of the
battery, and $\hat{\sigma}_{B}(k)=\sum_{n}r_{n}|n\rangle \langle n|$, known
as the passive state~\cite{Pusz,Lenard}, dependent on the
eigenvalues of $\hat{\rho}_{B}(k)$ that arranged in descending order (i.e., $%
r_{n}\leq r_{n+1}$). In terms of the ergotropy, one can define another kind
of the power
\begin{equation}
\mathcal{P}=\frac{\mathcal{E}_{B}(k)}{kt}=g\varepsilon \left( \frac{\bar{n}%
_{k}-\mathrm{Tr}_{B}[\hat{\sigma}_{B}(k)\hat{n}]}{k\tau }\right) .
\end{equation}%
Note that $\mathcal{E}_{B}(k)\leq E(k)$ and hence $\mathcal{P}\leq P$, where
the equality holds for a pure state of $\hat{\rho}_{B}(k)$. As shown in Fig.~%
\ref{fig5}(a) and (b), we show numerical results of $\mathcal{P}$ for $N_{A}=1$ and $4$ by taking fixed charging times $\tau =0.3$
(the open squares), $0.1$ (the crosses), and $0.01$ (the circles). With the
short-time case (i.e., $\tau =0.01$), one can see that the circles
show a good agreement with the analytical results of Eq.~(\ref{pcoh}). To
understand it, we calculate the purity of the battery
state $\varrho =\mathrm{Tr}_{B}[\hat{\rho}_{B}^{2}(k)]$, where $\varrho =1$
for a pure state and $\varrho <1$ for a mixed state. The purity $\varrho$ has also been investigated in the ultra-strong atom-field interaction~\cite{Shaghaghi} to show the pure state of the battery. As depicted in Fig.~\ref{fig5}(c) and (d), one
can see $\varrho \approx 1$ for $\tau=0.01$, indicating a nearly
pure state of $\hat{\rho}_{B}(k)$, which in turn gives $\mathcal{E}_{B}(k)\approx E(k)$
and hence $\mathcal{P}\approx P$. For a larger charging time $\tau =0.3$,
the battery state becomes more and more mixed, leading to a departure of $%
\mathcal{P}$ from $P$.

\section{Discussion and Conclusion}

We have generalized the repeated atom-battery interaction model (i.e., the so-called collisional
battery~\cite{Seah}) from the spin-$1/2$ charger to the case of a large spin $j=N_A/2$, where the battery is modeled by the
energy ladder with a finite number of levels $N_{B}+1$. Assuming little population over the
battery states $|0\rangle $ and $|N_{B}\rangle$ (corresponding to the no-boundary problem), we first derive a recursion relation of
the averaged excitation that stored in the battery (see Eq.~(\ref{nk}) and the Appendix A), which is
independent from any specific form of the atom state, and also free from the
initial state of $\hat{\rho}_{B}(0)$. Similar to the single-atom case (i.e.,
the number of two-level atoms $N_{A}=1$)~\cite{Seah}, the
incoherent and the coherent charging protocols have been investigated by considering the atoms prepared
in a mixed state and a coherent spin state $|\theta _{0},\phi _{0}\rangle $,
respectively. For the coherent protocol, the atomic coherence $\langle \hat{J%
}_{-}\rangle =j\sin (\theta _{0})\neq 0$, leading to a reduced charging time $k_{\mathrm{est}}\tau\sim N_{B}/N_{A}$, where $k_{\mathrm{est}}$ is the
number of collisions that for the battery being fully charged and $\tau$ is
the charging time per collision.

Next, we focus on the coherent charging process over the initially empty
battery in the short-time limit (i.e., $\tau \rightarrow 0$). Analytical
results of the average number of excitations [see Eqs.~(\ref{nk_ana}) and (%
\ref{nk_integ})] have been derived, which are related to the first-order
Bessel function of the first kind. Our results show that maximum of the
charging power $P_{\mathrm{coh,\max }}\approx 0.85g\varepsilon N_{A}$ can be
obtained by using the optimal coherent state $|\theta _{0},0\rangle $ with $%
\theta _{0}=\pi /2$. With a fixed charging time $k\tau =60/N_{A}$, we calculate the power against $\theta_0$ and $\tau$. For $\tau\ll 1$, the optimal state corresponds to $\theta _{0}=\pi /2$. When the atom state is imperfect and $\theta _{0}<\pi /2$, the power almost follows $P_{\mathrm{coh,\max }}$, provided that a relatively larger value of $\tau\sim 1.17(1-2\theta_0/\pi)$ is adopted.
As $\theta _{0}\rightarrow 0$, the maximum power $P_{\mathrm{inc,\max }}\approx 0.72g\varepsilon N_{A}$ appears at $\tau =1.17$, coincident with the single-atom case~\cite{Seah}.
Finally, another kind of the charging power has been investigated in terms of the so-called ergotropy. In the short-time
limit, we find that numerical results of the power show good agreement with
its analytical result, due to a nearly pure state of the battery in the early
charging steps (i.e., $k\lesssim 0.1k_{\mathrm{est}}$).

In summary, we have generalized the collision model to investigate coherent
charging of a single quantum battery by repeated interactions with finite $N_A$
two-level atoms. Analytical results of the average number of excitations and
hence the charging power have been derived in the short-time limit. Using an
optimal coherent spin state with $\theta_0=\pi /2$, we obtain the total charging time $k_{\mathrm{est}}\tau \sim N_{B}/N_{A}$ and the achievable charging power $0.85g\varepsilon N_{A}$, where $N_B$ is the number of the levels of the battery and $g\varepsilon N_{A}$ is the upper bound of the charging power. The faster charging time and the increased
amount of the power in comparison with the incoherent charging protocol uniquely arise from the phase coherence of the atoms. With a fixed charging time $k\tau$, we investigate the optimal choices of the initial atom state $\theta_0$ and the charging time per collision $\tau$. When $\theta_0$ largely departures from its optimal value $\pi/2$, the achievable charging power can almost maintain by choosing a relative large value of $\tau\sim 1.17(1-2\theta_0/\pi)$. Finally, we show that the
charging power defined by the ergotropy almost follows its analytical result, since the purity of the battery is almost equal to $1$, indicating a nearly pure state of the battery in the short-time limit. The above results rely on the assumption that the atom states are identical
for all the collisions. Indeed, it is interesting to investigate the dependence of the atom states on the charging steps (e.g., a defect state randomly appeared at one of the steps).

\begin{acknowledgments}
This work has been supported by the National Natural Science Foundation of
China (Nos. 12075209, 12005189, 11975205), and the Science Foundation of
Zhejiang Sci-Tech University (No. 18062145-Y).
\end{acknowledgments}

\begin{appendix}

\section{Details of Eqs.~(\ref{rho0}) and (\ref{nk})}

In the short-time limit, we have $\hat{U}_{\tau }\approx 1-i\tau (\hat{J}_{+}%
\hat{B}+\hat{J}_{-}\hat{B}^{\dag })$ and therefore Eq.~(\ref{rhobk}) becomes
\begin{eqnarray*}
\hat{\rho}_{B}(k) &=&\mathrm{Tr}_{A}[\hat{U}_{\tau }\hat{\rho}%
_{B}(k-1)\otimes \rho _{A}\hat{U}_{\tau }^{\dag }] \\
&\approx &\hat{\rho}_{B}(k-1)+\left[( \alpha ^{\ast }\hat{B}-\alpha \hat{B}%
^{\dag }) \hat{\rho}_{B}(k-1)+H.c.\right] \\
&\approx &(1+\alpha ^{\ast }\hat{B}-\alpha \hat{B}^{\dag })\hat{\rho}%
_{B}(k-1)(1-\alpha ^{\ast }\hat{B}+\alpha \hat{B}^{\dag }),
\end{eqnarray*}
with $\alpha =i\tau \langle \hat{J}_{-}\rangle $ and $\langle (\cdots
)\rangle =\mathrm{Tr}_{A}[\hat{\rho}_{A}(\cdots )]$. Iterating the above
equation for $k$ times, we obtain
\begin{eqnarray}
\hat{\rho}_{B}(k) &\approx &(1+\alpha ^{\ast }\hat{B}-\alpha \hat{B}^{\dag
})^{k}\rho _{B}(0)(1-\alpha ^{\ast }\hat{B}+\alpha \hat{B}^{\dag })^{k}
\notag \\
&\approx &\hat{D}^{\dag }(k\alpha )\hat{\rho}_{B}(0)\hat{D}(k\alpha ),
\end{eqnarray}%
as Eq.~(\ref{rho0}) in main text.

Next, we calculate the mean number of the excitations
\begin{eqnarray}
\bar{n}_{k} &=&\mathrm{Tr}_{B}[\hat{\rho}_{B}(k)\hat{n}]  \notag \\
&=&\mathrm{Tr}\left[ \hat{U}_{\tau }\hat{\rho}_{B}(k-1)\otimes \hat{\rho}_{A}%
\hat{U}_{\tau }^{\dag }\hat{n}\right]  \notag \\
&=&\mathrm{Tr}\left[ \hat{\rho}_{B}(k-1)\otimes \hat{\rho}_{A}\hat{U}_{\tau
}^{\dag }\hat{n}\hat{U}_{\tau }\right] ,  \label{nbar}
\end{eqnarray}%
where, in the second step, we have used Eq.~(\ref{rhobk}) in main text, and $%
\hat{n}=\sum_{n=0}^{N_{B}}n|n\rangle \langle n|$, satisfying
\begin{equation}
\lbrack \hat{B},\hat{n}]=\hat{B},\text{\ \ \ \ }~[\hat{B}^{\dag },\hat{n}]=-%
\hat{B^{\dag }}.  \label{bn}
\end{equation}%
Therefore, one can expand the term $\hat{U}_{\tau }^{\dag }\hat{n}\hat{U}%
_{\tau }$ using the Baker-Campbell-Hausdorff formula,
\begin{equation}
\hat{U}_{\tau }^{\dag }\hat{n}\hat{U}_{\tau }=e^{i\tau (\hat{J}_{+}\hat{B}+%
\hat{J}_{-}\hat{B}^{\dag })}\hat{n}e^{-i\tau (\hat{J}_{+}\hat{B}+\hat{J}_{-}%
\hat{B}^{\dag })}=\sum_{k=0}^{\infty }\frac{1}{k!}\hat{C}_{k},  \label{UnU}
\end{equation}%
where $\hat{C}_{k+1}=i\tau \lbrack \hat{J}_{+}\hat{B}+\hat{J}_{-}\hat{B}%
^{\dag },\hat{C}_{k}]$. Starting from $\hat{C}_{0}=\hat{n}$, we obtain%
\begin{eqnarray}
\hat{C}_{1} &=&i\tau \lbrack \hat{J}_{+}\hat{B}+\hat{J}_{-}\hat{B}^{\dag },%
\hat{C}_{0}]  \notag \\
&=&i\tau (\hat{J}_{+}[\hat{B},\hat{n}]+\hat{J}_{-}[\hat{B}^{\dag },\hat{n}])
\notag \\
&=&i\tau (\hat{J}_{+}\hat{B}-\hat{J}_{-}\hat{B}^{\dag }),  \label{C1}
\end{eqnarray}%
where we have used Eq.~(\ref{bn}). Next, we obtain%
\begin{eqnarray}
\hat{C}_{2} &=&i\tau \lbrack \hat{J}_{+}\hat{B}+\hat{J}_{-}\hat{B}^{\dag },%
\hat{C}_{1}]=2\tau ^{2}[\hat{J}_{+}\hat{B},\hat{J}_{-}\hat{B}^{\dag }]
\notag \\
&=&2\tau ^{2}(\hat{J}_{+}[\hat{B},\hat{J}_{-}\hat{B}^{\dag }]+[\hat{J}_{+},%
\hat{J}_{-}\hat{B}^{\dag }]\hat{B})\approx (2\tau )^{2}\hat{J}_{z},
\end{eqnarray}%
where we have used the commutation relation $[\hat{J}_{+},\hat{J}_{-}]=2\hat{%
J}_{z}$, as well as $[\hat{B},\hat{B}^{\dag }]\approx 0$ and $\hat{B}^{\dag }%
\hat{B}=1-|0\rangle\langle0|\approx 1$, valid for the no-boundary condition
(i.e., the occupations of $|0\rangle $ and $|N_{B}\rangle $ being
vanishing). Similarly, we obtain
\begin{eqnarray*}
\hat{C}_{3} &=&i\tau \lbrack \hat{J}_{+}\hat{B}+\hat{J}_{-}\hat{B}^{\dag },%
\hat{C}_{2}]=-4i\tau ^{3}(\hat{J}_{+}\hat{B}-\hat{J}_{-}\hat{B}^{\dag }), \\
\hat{C}_{4} &=&i\tau \lbrack \hat{J}_{+}\hat{B}+\hat{J}_{-}\hat{B}^{\dag },%
\hat{C}_{3}]\approx -\left( 2\tau \right) ^{4}\hat{J}_{z},
\end{eqnarray*}%
and so on. Finally, one can easily obtain
\begin{eqnarray}
\hat{U}_{\tau }^{\dag }\hat{n}\hat{U}_{\tau } &\approx &\hat{n}+\frac{\left(
\hat{J}_{-}\hat{B}^{\dag }-\hat{J}_{+}\hat{B}\right) }{2i}\sum_{k=0}^{\infty
}\frac{\left( -1\right) ^{k}}{\left( 2k+1\right) !}\left( 2\tau \right)
^{2k+1}  \notag \\
&&+\hat{J}_{z}\sum_{k=0}^{\infty }\frac{\left( -1\right) ^{k}}{\left(
2k+2\right) !}\left( 2\tau \right) ^{2k+2}  \notag \\
&=&\hat{n}+\sin \left( 2\tau \right) \frac{\left( \hat{J}_{-}\hat{B}^{\dag }-%
\hat{J}_{+}\hat{B}\right) }{2i}+2\sin ^{2}\left( \tau \right) \hat{J}_{z}.
\notag  \label{evo n}
\end{eqnarray}%
Therefore, Eq.~(\ref{nbar}) becomes
\begin{equation}
\bar{n}_{k}\approx \bar{n}_{k-1}+2\sin ^{2}(\tau )\langle \hat{J}_{z}\rangle
+\sin (2\tau )\mathrm{Im}\left(\langle \hat{J}_{-}\rangle \beta _{k-1}^{\ast
}\right),
\end{equation}
as Eq.~(\ref{nk}) in main text, where $\beta _{k}=\mathrm{Tr}_{B}[\hat{\rho}%
_{B}(k)\hat{B}]$.

\section{Details of Eqs.~(\ref{beta2}) and (\ref{nk_ana})}

First, we calculate $\beta _{k}$ for the battery state $\hat{\rho}_{B}(k)$
defined by Eq.(\ref{rhobk}) in main text,
\begin{eqnarray}
\beta _{k} &=&\mathrm{Tr}_{B}[\hat{\rho}_{B}(k)\hat{B}]  \notag \\
&=&\mathrm{Tr}[\hat{U}_{\tau }\hat{\rho}_{B}(k-1)\otimes \hat{\rho}_{A}\hat{U%
}_{\tau }^{\dag }\hat{B}]  \notag \\
&=&\mathrm{Tr}[\hat{\rho}_{B}(k-1)\otimes \hat{\rho}_{A}\hat{U}_{\tau
}^{\dag }\hat{B}\hat{U}_{\tau }].
\end{eqnarray}%
Similar to Eq.~(\ref{UnU}), we deal with the term
\begin{equation}
\hat{U}_{\tau }^{\dag }\hat{B}\hat{U}_{\tau }=e^{i\tau (\hat{J}_{+}\hat{B}+%
\hat{J}_{-}\hat{B}^{\dag })}\hat{B}e^{-i\tau (\hat{J}_{+}\hat{B}+\hat{J}_{-}%
\hat{B}^{\dag })}=\sum_{k=0}^{\infty }\frac{1}{k!}\hat{D}_{k},  \label{UBU}
\end{equation}%
where $\hat{D}_{k+1}=i\tau \lbrack \hat{J}_{+}\hat{B}+\hat{J}_{-}\hat{B}%
^{\dag },\hat{D}_{k}]$, with $\hat{D}_{0}=\hat{B}$. In the short-time limit,
the first-order expansion is enough, i.e.,%
\begin{equation}
\hat{U}_{\tau }^{\dag }\hat{B}\hat{U}_{\tau }\approx \hat{B}+\hat{D}_{1},
\end{equation}%
where
\begin{eqnarray}
\hat{D}_{1} &=&i\tau \lbrack \hat{J}_{+}\hat{B}+\hat{J}_{-}\hat{B}^{\dag },%
\hat{D}_{0}]  \notag \\
&=&i\tau \hat{J}_{-}[\hat{B}^{\dag },\hat{B}]=i\tau \hat{J}_{-}\left(
|N_{B}\rangle \langle N_{B}|-|0\rangle \langle 0|\right) .
\end{eqnarray}%
Therefore, we obtain
\begin{eqnarray}
\beta _{k} &\approx &\mathrm{Tr}_{B}\left[ \hat{\rho}_{B}(k-1)\hat{B}+\alpha
\hat{\rho}_{B}(k-1)\hat{D}_{1}\right]  \notag \\
&=&\beta _{k-1}+\alpha \mathrm{Tr}_{B}\left[ \hat{\rho}_{B}(k-1)\left(
|N_{B}\rangle \langle N_{B}|-|0\rangle \langle 0|\right) \right]  \notag \\
&=&\beta _{k-1}+\alpha \left( \langle N_{B}|\hat{\rho}_{B}(k-1)|N_{B}\rangle
-\langle 0|\hat{\rho}_{B}(k-1)|0\rangle \right)  \notag \\
&=&\beta _{0}+\sum_{k^{\prime }=0}^{k-1}\alpha \left( \langle N_{B}|\hat{\rho%
}_{B}(k^{\prime })|N_{B}\rangle -\langle 0|\hat{\rho}_{B}(k^{\prime
})|0\rangle \right) ,
\end{eqnarray}%
where the last second result has been iterated for $k$ times. With the
no-boundary condition, we simply obtain $\beta _{k}\approx \beta _{0}$;
Neglecting only the upper boundary $|N_{B}\rangle $, we obtain Eqs.~(\ref%
{beta2}) and (\ref{beta_k}) in main text, where the lower boundary $%
|0\rangle $ has been taken into accounted.

Next, we analysis $\beta _{k}$ for an arbitrary state $\hat{\rho}%
_{B}(k)=\sum_{i}p_{i}|\psi _{B}^{(i)}\rangle \langle \psi _{B}^{(i)}|$,
where $|\psi _{B}^{(i)}\rangle =\sum_{n=0}^{N_{B}}c_{n}^{(i)}|n\rangle $ and
$\sum_{i}p_{i}=1$, which gives the Cauchy-Schwartz inequality:
\begin{eqnarray}
|\beta _{k}| &=&\left\vert \sum_{n=0}^{N_{B}}\left\langle n\right\vert \hat{%
\rho}_{B}(k)\hat{B}\left\vert n\right\rangle \right\vert
\leq\sum_{i}p_{i}\left\vert
\sum_{n=1}^{N_{B}}c_{n}^{(i)}c_{n-1}^{(i)\ast}\right\vert  \notag \\
&\leq &\sum_{i}p_{i}\sqrt{\sum_{n=1}^{N_{B}}\left|c_{n}^{(i)}\right|^{2}%
\sum_{n^{\prime}=1}^{N_{B}}\left|c_{n^{\prime }-1}^{(i)}\right|^{2}}  \notag
\\
&=&\sum_{i}p_{i}\sqrt{\left(1-\left|c_{0}^{(i)}\right|^{2}\right)\left(1-%
\left|c_{N_{B}}^{(i)}\right|^{2}\right)}  \notag \\
&\leq &1.  \notag
\end{eqnarray}%
Note that the equality in the second step holds for $c_{n}^{(i)}c_{n-1}^{(i)%
\ast}\in\mathbb{R}$. The following two equalities hold for $%
c_{n}^{(i)}=c_{n-1}^{(i)\ast }$ and $c_{0}^{(i)}=c_{N_{B}}^{(i)}=0$,
respectively. Therefore, it is easy to obtain $|\beta _{k}|\in \lbrack 0,1)$%
, which has been used in Eq.~(\ref{n bound}).

Finally, we calculate Eq.~(\ref{nbar}) in the short-time limit to derive
Eq.~(\ref{nk_ana}) in main text. Using $\hat{U}_{\tau }^{\dag }\hat{n}\hat{U}%
_{\tau }\approx \hat{n}+\hat{C}_{1}=\hat{n}+i\tau (\hat{J}_{+}\hat{B}-\hat{J}%
_{-}\hat{B}^{\dag })$, we obtain%
\begin{eqnarray}
\bar{n}_{k} &\approx &\mathrm{Tr}_{B}\left[ \hat{\rho}_{B}(k-1)\hat{n}-\hat{%
\rho}_{B}(k-1)(\alpha \hat{B}^{\dag }+\alpha ^{\ast }\hat{B})\right]   \notag
\\
&=&\bar{n}_{k-1}-(\alpha \beta _{k-1}^{\ast }+\alpha ^{\ast }\beta _{k-1}),
\end{eqnarray}%
as Eq.~(\ref{nklimit}) in main text. Iterating the above equation for $k$
times, we further obtain
\begin{equation}
\bar{n}_{k}\approx \bar{n}_{0}-\sum_{k^{\prime }=0}^{k-1}(\alpha \beta
_{k^{\prime }}^{\ast }+\alpha ^{\ast }\beta _{k^{\prime }}),
\end{equation}%
where $\bar{n}_{0}=0$ for $\hat{\rho}_{B}(0)=|0\rangle \langle 0|$, and $%
\beta _{k}$ depends on the terms $\langle 0|(\alpha \hat{B}^{\dag }-\alpha
^{\ast }\hat{B})^{l}|0\rangle $, as Eq.~(\ref{0D0}) in main text. For odd $l$%
, e.g., $l=1$, it is easy to see $\langle 0|(\alpha \hat{B}^{\dag }-\alpha
^{\ast }\hat{B})|0\rangle =\alpha \langle 0|1\rangle =0$. As a result, we
only consider even $l=2n$, e.g., $n=1$,
\begin{eqnarray*}
&&\langle 0|(\alpha \hat{B}^{\dag }-\alpha ^{\ast }\hat{B})^{2}|0\rangle  \\
&=&\langle 0|\left[ (\alpha \hat{B}^{\dag })^{2}+(-\alpha ^{\ast }\hat{B}%
)^{2}-|\alpha |^{2}\hat{B}^{\dag }\hat{B}-|\alpha |^{2}\hat{B}\hat{B}^{\dag }%
\right] |0\rangle  \\
&=&-|\alpha |^{2}\langle 0|\left[ \hat{B}^{\dag }\hat{B}+\hat{B}\hat{B}%
^{\dag }\right] |0\rangle =-|\alpha |^{2}\langle 0|\hat{B}\hat{B}^{\dag
}|0\rangle  \\
&=&-|\alpha |^{2}.
\end{eqnarray*}%
To obtain $\langle 0|(\alpha \hat{B}^{\dag }-\alpha ^{\ast }\hat{B}%
)^{2n}|0\rangle \neq 0$, it requires the numbers of $\hat{B}$ and $\hat{B}%
^{\dag }$ are equal. Furthermore, the ordering over the ladder operators $%
\hat{B}$ and $\hat{B}^{\dag }$ corresponds to the evolution paths from the
\textquotedblleft initial" state $|0\rangle $ to the \textquotedblleft
final" state $|0\rangle $, as depicted by Fig.~\ref{fig3} (a), where each
path gives the same value $(-|\alpha |^{2})^{n}$ and the number of all
possible paths is given by the Catalan number $C_{n}=\frac{1}{n+1}\binom{2n}{%
n}$. Therefore, we obtain%
\begin{equation}
\langle 0|(\alpha \hat{B}^{\dag }-\alpha ^{\ast }\hat{B})^{2n}|0\rangle
=(-1)^{n}|\alpha |^{2n}C_{n}.
\end{equation}%
Substituting it into Eq.~(%
\ref{0D0}), we obtain Eqs.~(\ref{ka00}) and~(\ref{beta_k}) in main text, and
therefore
\begin{eqnarray}
\bar{n}_{k} &\approx &2\sum_{k^{\prime }=0}^{k-1}\sum_{l=0}^{k^{\prime
}-1}\left( \frac{J_{1}(2l|\alpha |)}{l}\right) ^{2}=2\sum_{k^{\prime
}=1}^{k-1}\sum_{l=0}^{k^{\prime }-1}\left( \frac{J_{1}(2l|\alpha |)}{l}%
\right) ^{2}  \notag \\
&=&2\sum_{l=0}^{k-2}\sum_{k^{\prime }=l+1}^{k-1}\left( \frac{J_{1}(2l|\alpha
|)}{l}\right) ^{2}  \notag \\
&=&2\sum_{l=0}^{k-2}(k-1-l)\left( \frac{J_{1}(2l|\alpha |)}{l}\right) ^{2},
\end{eqnarray}%
where, in the last two steps, we have interchanged the order of summation $%
\sum_{k^{\prime }=1}^{k-1}\sum_{l=0}^{k^{\prime }-1}(\cdots
)=\sum_{l=0}^{k-2}\sum_{k^{\prime }=l+1}^{k-1}(\cdots )$, and note the inner
sum $\sum_{k^{\prime }=l+1}^{k-1}(\cdots )=(k-l-1)(\cdots )$.

\end{appendix}

\end{document}